\documentclass[journal=jcisd8,manuscript=article]{achemso}
\setkeys{acs}{doi = true}

\usepackage[version=3]{mhchem} 
\usepackage{siunitx}
\usepackage{amssymb}
\usepackage{glossaries}
\usepackage{multirow}
\usepackage[dvipsnames]{xcolor}
\usepackage{soul,color}

\author{Christopher Kuenneth}
\affiliation[LANL-MST]{Materials Science and Technology Division, Los Alamos National Laboratory, Los Alamos, NM 87545, USA}
\alsoaffiliation[Gatech]{School of Materials Science and Engineering, Georgia Institute of Technology, Atlanta, Georgia 30332, USA}

\author{Jessica Lalonde}
\affiliation[LANL-bio]{Materials Science and Technology Division, Los Alamos National Laboratory, Los Alamos, NM 87545, USA}
\alsoaffiliation[Duke]{Department of Mechanical Engineering and Materials Science, Duke University, Durham, North Carolina 27708, USA}

\author{Babetta L. Marrone}
\affiliation[LANL-bio]{Bioscience Division, Los Alamos National Laboratory, Los Alamos, NM 87545, USA}

\author{Carl N. Iverson}
\affiliation[LANL-CHEM]{Chemistry Division, Los Alamos National Laboratory, Los Alamos, NM 87545, USA}

\author{Rampi Ramprasad}
\affiliation[Gatech]{School of Materials Science and Engineering, Georgia Institute of Technology, Atlanta, Georgia 30332, USA}

\author{Ghanshyam Pilania}
\email{gpilania@lanl.gov}
\affiliation[LANL-MST]{Materials Science and Technology Division, Los Alamos National Laboratory, Los Alamos, NM 87545, USA}

\title[]
  {Bioplastic Design using Multitask Deep Neural Networks}

\begin{document}







\begin{abstract} 
Non-degradable plastic waste stays for decades on land and in water, jeopardizing our environment; yet our modern lifestyle and current technologies are impossible to sustain without plastics. Bio-synthesized and biodegradable alternatives such as the polymer family of polyhydroxyalkanoates (PHAs) have the potential to replace large portions of the world's plastic supply with \textit{cradle-to-cradle} materials, but their chemical complexity and diversity limit traditional resource-intensive experimentation. In this work, we develop multitask deep neural network property predictors using available experimental data for a diverse set of nearly $23\,000$ homo- and copolymer chemistries. Using the predictors, we identify 14 PHA-based bioplastics from a search space of almost 1.4 million candidates which could serve as potential replacements for seven petroleum-based commodity plastics that account for \SI{75}{\%} of the world's yearly plastic production. We discuss possible synthesis routes for these identified promising materials. The developed multitask polymer property predictors are made available as a part of the Polymer Genome project at \url{https://PolymerGenome.org}.
\end{abstract}

\section{Introduction}

Plastics are an integral part of our everyday life and modern technology. Their simple, yet diverse, chemistries and tunable properties make plastics versatile and desirable; plastics display high or low flexibility, strength, thermal, or electronic conductivity along with low cost, low weight, and abundance\cite{Satti_2020}. The global plastic production of 2019 amounts to an unimaginable 368 million tonnes and is expected to further increase in the coming years. About \SI{40}{\%} (145 million tonnes) of the yearly plastic production accounts for packaging products such as bags, food containers, cutlery, or bottles, which have very short service lifetimes and often end up in landfills, seawater, or other natural environments\cite{plasticseurop,Naser2021,Geyer2017}. It is therefore not surprising that packaging plastics are one of the largest polluters of our world's ecosystems, severely threatening the existence of animals and humans through waste and microplastic particles on land and in oceans that last for decades or longer\cite{Lim2021}. Finding eco-friendly plastics (bioplastics) with properties akin to conventional plastics but with sustainable recycling options is therefore of utmost importance for a circular economy.\cite{Satti_2020} The bio-derived and biodegradable family of polyhydroxyalkanoates (PHAs) is a promising cradle-to-cradle material that can be synthesized by several microorganisms directly using sunlight and CO$_2$ from the environment or industrial point sources.\cite{Naser2021,Poltronieri_2019,goPHA} Diverse chemistries harbored in PHAs span a large property space with ample opportunities to design mechanical and thermal properties such as the Young's modulus ($E$), tensile strength ($\sigma$), elongation ($\epsilon$), glass transition temperature ($T_\text{g}$), melting temperature ($T_\text{m}$), and degradation temperature ($T_\text{d}$) \cite{Follain_2014, Naser2021,pilania2019machine,bejagam2020molecular,bejagam2021composition,bejagam2022predicting,bejagam2022machine}.

\begin{figure}[hbt]
 \includegraphics[width=1\textwidth]{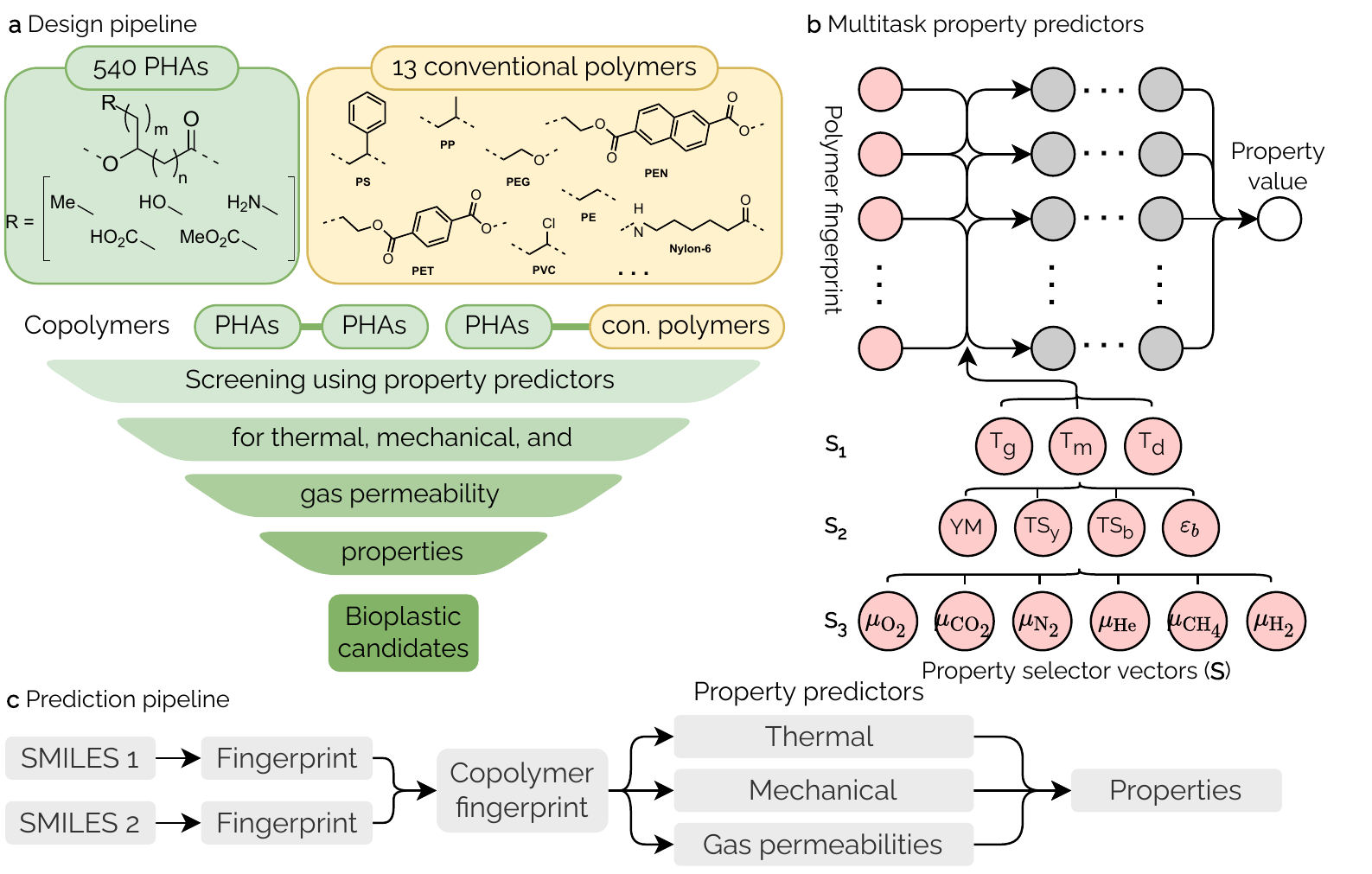}
  \caption{Bioplastic design using multitask deep learning predictors. \textbf{a} Design pipeline. A large search space is created by combining 540 polyhydroxyalkanoates (PHAs) and 13 conventional polymers to copolymers. Property predictors and property requirements of commonly used polymers allow us to identify bioplastic candidates within the search space. \textbf{b} Architecture of the multitask neural network predictors. Three separate predictors are trained; one for each of the selector vectors ($\mathbf{S}_1$, $\mathbf{S}_2$, $\mathbf{S}_3$). \textbf{c} Prediction pipeline. The two SMILES\cite{Weininger1988} strings belong to comonomers in a copolymer.}
  \label{fig:pha_design}
\end{figure}

PHAs provide copious opportunities for chemical modification and property modulation \cite{Sharma_2021, Naser2021}. Key parameters of these modifications are the numbers of carbons in the main-chain and side-chain ($n$ and $m$ in Figure \ref{fig:pha_design}a), and the terminating functional groups of the side-chain ($R$ in Figure \ref{fig:pha_design}a) \cite{Sharma_2021, Pryadko_2021}. For instance, the most widely known PHA, poly-3-hydroxybutyrate (P3HB, with $n = 1$, $m = 1$, and no $R$), is brittle and inflexible\cite{McAdam2020}. As the number of carbon atoms in the backbone increases, the resulting polymers tend to display higher elongation at break ($\epsilon_\text{b}$) combined with improved mechanical strength and enhanced tendency for degradability \cite{Naser2021}. Also, PHAs with side-chain-terminating phenyl groups exhibit higher $T_\text{g}$s because of increased rigidity due to enhanced interchain interactions resulting from the polar side chain functional groups\cite{Jiang_2020, Sharma_2021}. Besides systematic structural and chemical alterations, copolymers provide an additional knob to grow the accessible property space by not only combining multiple PHA-based motifs but also PHAs with conventional polymers \cite{Winnacker2017}. In the past, PHA-PHA copolymers have been found to improve mechanical properties while keeping high $T_\text{m}$ and low $T_\text{g}$ values, which is ideal for applications that require large temperature operation windows \cite{Naser2021, bejagam2022machine}. By forming copolymers of PHAs with other conventional polymers, one may harness synergistic effects, potentially leading to improved recyclability and enhanced mechanical strength. 

Just contemplating copolymer compositions on a rather coarse composition grid ($ c = 0, 0.1, \dots, 1 $), the total number of distinct PHA-PHA copolymer possibilities is far beyond a million, effectively rendering trial and error-based high-throughput experiments an impractical route of searching for application-specific candidate materials\cite{Albright_2021}. Also, the sheer size of the search space disqualifies time-consuming computational methods such as density functional theory (DFT) or even classical molecular dynamics (MD) simulations. The burgeoning field of polymer informatics\cite{Chen2021b,Batra2020,Ramprasad2017,Audus2017,Peerless2019,Adams2008} offers an exciting alternative route to address such search problems by using modern data-driven machine learning approaches \cite{Jiang2020,zhong2021machine}.

The present study, with the details of the workflow and machine learning framework outlined in Figure \ref{fig:pha_design}, has several vital elements. First, we develop efficient multitask deep neural network-based multiproperty predictors for copolymers that forecast three different thermal  ($T_\text{g}$, $T_\text{m}$, and $T_\text{d}$), four different mechanical ($E$, $\sigma_\text{y}$, $\sigma_\text{b}$, and $\epsilon_\text{b}$), and six gas permeability ($\mu_{g} | g \in \{\text{O}_2, \text{CO}_2, \text{N}_2, \text{H}_2, \text{He}, \text{CH}_4 \}$) properties using nearly $23\,000$ experimental data points pertaining to a diverse range of homo- and copolymer chemistries. Here, $T_\text{g}$, $T_\text{m}$, $T_\text{d}$, $E$, $\sigma_\text{y}$, $\sigma_\text{b}$, and $\epsilon_\text{b}$ are the glass transition temperature, melting temperature, degradation temperature, Young's modulus, tensile strength at yield, tensile strength at break, and elongation at break, respectively. $\mu_{\text{O}_2}$, $\mu_{\text{CO}_2}$, $\mu_{\text{N}_2}$, $\mu_{\text{H}_2}$, $\mu_{\text{He}}$, and $\mu_{\text{CH}_4}$ stand for the gas permeabilities of $\text{O}_2$, $\text{CO}_2$, $\text{N}_2$, $\text{H}_2$, $\text{He}$, and $\text{CH}_4$. The thermal, mechanical, and gas permeability properties are selected as they play a critical role in the design and selection of plastics for packaging and other large-scale industrial applications. Second, we create a bioplastic candidate space of nearly 1.4 million bioplastics, which is spanned by 540 PHAs and 13 conventional polymer chemistries. Third, we follow a two-step protocol to find several PHA-only and PHA-conventional polymer bio-replacements in the candidate space for seven petroleum-based and commonly used plastics. Possible synthesis routes of the bio-replacements are discussed. This work represents the state-of-the-art in polymer informatics, and contributes to and accelerates the identification of sustainable functional polymer candidate materials. 

\section{Results and Discussion}

\begin{table}[hbtp]
  \caption{An overview of our data set used for training the multitask predictors. The property portfolio for the three subgroups, associated property ranges, units of measurements as well as the number of homo- and copolymers included in the dataset are outlined.}
  \label{tbl:data_set}
\resizebox{\textwidth}{!}{%
\begin{tabular}{lccrrc|r}

 & Symbol & Unit & Homopolymer & Copolymer & Range & Total\\
\hline
\multicolumn{5}{l}{Thermal properties} & \\
\hline
Glass transition temp. & $T_\text{g}$ & K &  $5\,183$ &  $3\,312$ & $[80, 873]$ & $8\,495$ \\
Melting temp. & $T_\text{m}$ & K & $2\,132$ & $1\,523$ & $[215, 860]$ & $3\,655$ \\
Degradation temp. & $T_\text{d}$  &  K & $3\,584$ &  $1\,064$ & $[291, 1\,173]$ &  $4\,648$ \\
\hline
\multicolumn{5}{l}{Mechanical properties} & \\
\hline  
Young's modulus & $E$  & MPa &  592 &  322 & $[0.2, 4000]$ & 914 \\
Tensile strength at yield & $\sigma_\text{y}$ & MPa &  216 &  78 & $[0.01, 132]$ & 294 \\
Tensile strength at break & $\sigma_\text{b}$  & MPa &  663 &  318 & $[0.04, 200]$ & 981 \\
Elongation at break & $\epsilon_\text{b}$  &   &  868 &  260 & $[0.3, 995]$ & $1\,128$ \\
\hline
\multicolumn{5}{l}{Gas permeability properties} & \\
\hline
$\text{O}_2$ & $\mu_{\text{O}_2}$  & barrer &  420 &  210 & $[5\cdot10^{-6}, 1\,000]$ & 630 \\
$\text{CO}_2$ & $\mu_{\text{CO}_2}$  & barrer & 313  &  119 & $[10^{-6}, 4\,756]$ & 432 \\
$\text{N}_2$ & $\mu_{\text{N}_2}$  & barrer & 417  &  99 & $[3\cdot 10^{-5}, 480]$ & 516 \\
$\text{H}_2$ & $\mu_{\text{H}_2}$  & barrer & 266  &  46 & $[2\cdot 10^{-2}, 5\,000]$ & 312 \\
$\text{He}$ & $\mu_{\text{He}}$  & barrer & 261  &  58 & $[5\cdot 10^{-2}, 1\,950]$ & 319 \\
$\text{CH}_4$ & $\mu_{\text{CH}_4}$  & barrer & 360  &  47 & $[4\cdot 10^{-4}, 1\,690]$ & 407 \\
\hline
Total & & & $15\,344$ & $7\,512$  & &  $22\,856$ \\
\end{tabular}
}
\end{table}

\paragraph{Data Set} Our data set for training of the multitask property predictors includes a total of $22\,856$ homopolymer (\SI{\approx60}{\%}) and copolymer (\SI{\approx30}{\%}) data points of the thermal, mechanical, and the small molecule gas permeability properties as reported in Table \ref{tbl:data_set}. Each of the $7\,512$ copolymer data points involves two distinct comonomers at various compositions while spanning over $1\,440$ distinct copolymer chemistries. Homo- and copolymer data points of $T_\text{g}$, $T_\text{m}$, and $T_\text{d}$, and homopolymer data points of $\mu_g$s, $E$, and $\sigma_\text{b}$ were already utilized in previous studies\cite{Kim2018,Kim2019,Jha2019a,Kuenneth2021,Kuenneth2021a}. The copolymer data points belonging to $\mu_g$s, $E$, $\sigma_\text{y}$, $\sigma_\text{b}$, and $\epsilon_\text{b}$, and homopolymer data points of $\sigma_\text{y}$ and $\epsilon_\text{b}$ were collected from the PolyInfo\cite{polyinfo} repository for this study. For consistency and uniformity, only $T_\text{g}$ and $T_\text{m}$ data points measured via differential scanning calorimetry (DSC), $T_d$ data points measured via thermogravimetric analysis (TGA), and mechanical data points recorded around room temperature (\SI{300}{K}) were included in the data set. Moreover, for configurational consistency, all copolymer data points in this study are from random copolymers. As part of an additional curation step and our due diligence strategy, we employed a clustering algorithm (as implemented in Scikit-learn\cite{Varoquaux2015}) to identify outliers and select suspicious data points for manual inspection. The degree of polymerization and molecular weight were not taken into account because they were not uniformly available for all data points. Mandated by the multitask method, all property values were scaled to the range of $[0, 1]$ (min-max scaling) for training and transformed back to the actual ranges before computing the respective error metrics. Additionally, $\epsilon_\text{b}$ and the gas permeabilities were transformed to the log base 10 scale ($ x \mapsto \log_{10} (x + 1)$) before training because of their power-law-shaped data distributions (c.f., Supplementary Figures S4 and S5).

\paragraph{Property Predictors}

Multitask deep neural networks with meta learners have shown best-in-class performance in past polymer informatics studies\cite{Kuenneth2021,Kuenneth2021a} due to their ability to utilize inherent correlations in data that helps to overcome data sparsity. Here, we create three multiproperty predictors (one for each category in Table \ref{tbl:data_set}) to predict, in total, 13 polymer properties using the data set and categories profiled in Table \ref{tbl:data_set} and fingerprints outlined in the Methods section. Figure \ref{fig:pha_design}b schematically shows the architecture of the multitask predictors, while implementation details are given in the Methods section.

The developed meta learners display an outstanding overall coefficient of determination ($R^2$) value of 0.97, with all $R^2$s higher than 0.9. The root-mean-square error (RMSE) and $R^2$ values of all properties are reported in Table \ref{tbl:prop_metric}. The thermal property predictor performs very well with $R^2$s as high as 0.98, 0.97, and 0.96 for $T_\text{g}$, $T_\text{m}$, and $T_\text{d}$, respectively. This is expected because of the large number of data points and high data fidelity of the thermal data points. It should be noted that the reported values here are slightly better than those reported in Reference \citenum{Kuenneth2021a}, which uses a very similar thermal property data set. This is because of the extra data curation and cleaning steps adopted in this work, as discussed above in the Data Set section. The mechanical and gas permeability predictors show very high $R^2$s of 0.94, 0.96, 0.94, and 0.91 for $E$, $\sigma_y$, $\sigma_b$, and $\epsilon_\text{b}$, respectively, and 0.99, 0.99, 0.99, 1.00, 0.99, and 0.99 for the six gases $g \in \{\text{O}_2, \text{CO}_2, \text{N}_2, \text{H}_2, \text{He}, \text{CH}_4 \}$, respectively. The overall performance of the three developed predictors with averaged $R^2$s of 0.97, 0.94, and 0.99 is exceptional and may be credited to the large data set of almost $23\,000$ data points, additional data curation measures, well-conditioned and smooth fingerprints, and fully-hyperparameter-optimized multitask deep neural networks with meta learners. The individual parity plots of the meta learners for each property can be found in the Supplementary Figures S6-S8.

\begin{table}[hbtp]
  \caption{The RMSEs and $R^2$s averages of the five cross-validation models and meta learner from predictions on the respective validation data sets. The reported uncertainties are the \SI{68}{\%} confidence intervals ($2\sigma$).}
  \label{tbl:prop_metric} 
\begin{tabular}{llrrrr}
\hline
  Symbol$^a$   &   Unit & \multicolumn{2}{c}{Cross-validation} & \multicolumn{2}{c}{Meta learner} \\
                    &      &             RMSE &             $R^2$ &    RMSE & $R^2$ \\
\hline
\hline
\multicolumn{6}{l}{Thermal properties} \\
\hline
       $T_\text{g}$ &      K &   \num{29.78\pm1.26} & \num{0.92\pm0.01} &   13.04 &  0.98 \\
       $T_\text{m}$ &      K &   \num{40.17\pm0.83} & \num{0.84\pm0.01} &   16.67 &  0.97 \\
       $T_\text{d}$ &      K &   \num{62.16\pm2.52} & \num{0.72\pm0.02} &   23.84 &  0.96 \\
\hline
\multicolumn{6}{l}{Mechanical properties} \\
\hline
                $E$ &    MPa & \num{475.34\pm31.84} & \num{0.78\pm0.03} &  237.2 &  0.94 \\
  $\sigma_\text{y}$ &    MPa &   \num{15.43\pm3.81} & \num{0.79\pm0.12} &    7.1 &  0.96 \\
  $\sigma_\text{b}$ &    MPa &   \num{18.82\pm1.00} & \num{0.77\pm0.02} &   9.81 &  0.94 \\
$\epsilon_\text{b}$ $^b$ &        &    \num{0.43\pm0.04} & \num{0.59\pm0.10} &    0.2 &  0.91 \\
\hline
\multicolumn{6}{l}{Gas permeability properties} \\
\hline
 $\mu_{\text{O}_2}$ $^b$ & barrer & \num{0.13\pm0.02} & \num{0.97\pm0.02} &    0.07 &  0.99 \\
$\mu_{\text{CO}_2}$ $^b$ & barrer & \num{0.20\pm0.04} & \num{0.96\pm0.02} &    0.11 &  0.99 \\
 $\mu_{\text{N}_2}$ $^b$ & barrer & \num{0.12\pm0.04} & \num{0.96\pm0.03} &    0.05 &  0.99 \\
 $\mu_{\text{H}_2}$ $^b$ & barrer & \num{0.14\pm0.02} & \num{0.97\pm0.01} &    0.06 &   1.0 \\
  $\mu_{\text{He}}$ $^b$ & barrer & \num{0.14\pm0.01} & \num{0.96\pm0.01} &    0.06 &  0.99 \\
$\mu_{\text{CH}_4}$ $^b$ & barrer & \num{0.16\pm0.03} & \num{0.96\pm0.01} &    0.06 &  0.99 \\
\hline
\end{tabular}

\footnotesize{$^a$ See Table \ref{tbl:data_set} for symbol definition.}

\footnotesize{$^b$ Trained on log base 10 scale ($ x \mapsto \log_{10} (x + 1)$). RMSE and $R^2$ values are reported on this scale.}\\
\end{table}

\paragraph{Bioplastic Search Space}

In the next step, we consider a bioplastic space that can be searched using the property predictors developed in the last section. As shown in Figure \ref{fig:pha_design}a, 540 PHAs and 13 conventional polymers define and bound this space. The 540 PHAs are devised through variations of the number of carbon atoms in the main-chain and side-chain from 1 to 6 ($n$ and $m$ in Figure \ref{fig:pha_design}a), and by terminating the side-chains with 17 different functional groups (see Supplementary Figure S1). The bio-copolymers of this space are generated by the outer product of PHAs and conventional polymers at eleven different compositions ($ c = 0, 0.1, \dots, 1$). The total number of bioplastics in the search space amounts to $1\,373\,503$ and is composed of $553$ homopolymers, $146\,070$ copolymers of PHAs-only, and $7\,033$ copolymers of PHAs and conventional polymers. The 13 conventional polymers were selected according to the list of most commonly used plastics and are documented in Supplementary Figure S2.

\begin{figure}[hbt]
 \includegraphics{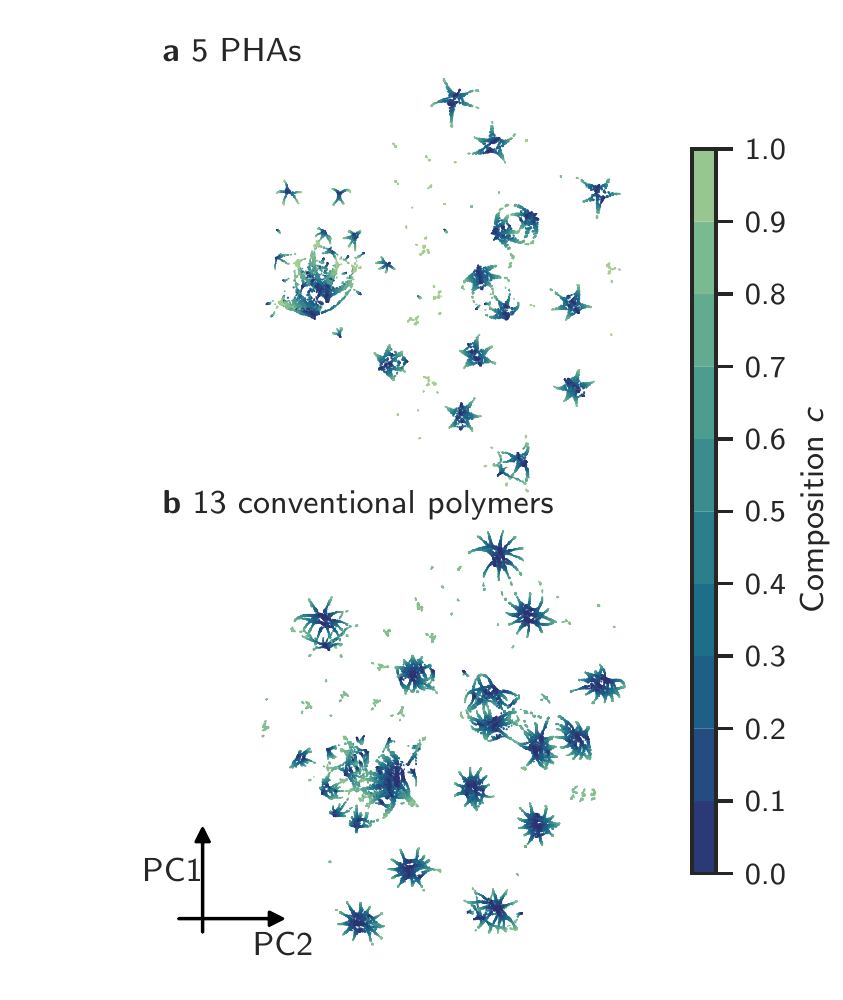}
  \caption{Two-dimensional UMAP\cite{mcinnes2018umap-software} plot of two fingerprint subspaces. \textbf{a} Five PHAs. The dark green dots ($c=1$) show the fingerprints of the SMILES strings \texttt{[*]OCCC(=O)[*]}, \texttt{[*]OC(O)CC(=O)[*]}, \texttt{[*]OC(C(=O)O)CC(=O)[*]}, \texttt{[*]OC(C(=O)OC)CC(=O)[*]}, and \texttt{[*]OC(N)CC(=O)[*]}. \textbf{b} 13 conventional polymers. The dark green dots show the fingerprints of the 13 conventional polymers. The dark blue dots ($c=0$) in the panels \textbf{a} and \textbf{b} indicate the fingerprints of the remaining 548 and 540 polymers in the bioplastic search space (a total of 553 polymers), respectively. The dots with intermediate colors (green to blue) indicate the fingerprints of connecting copolymers. PC1 and PC2 represent the first and second principal components in the UMAP projection, respectively.}
  \label{fig:umap}
\end{figure}

Figures \ref{fig:umap}a and \ref{fig:umap}b display the 2D uniform manifold approximations and projections (UMAPs)\cite{mcinnes2018umap-software} of two different fingerprint subspaces. The fingerprint subspace of Figure \ref{fig:umap}a contains five PHAs (green dots, $c=1$), the remaining 548 polymers (blue dots, $c=0$), and copolymers ($c=0.1, 0.2, \dots,0.9$) that connect the five PHAs and the remaining 548 polymers. Interestingly, the UMAP method has identified similar polymers and aggregated them into the shape of stars. The corner vertices of these stars show the fingerprints of the five PHAs, while all other dots in the stars indicate the fingerprints of polymers of similar chemistry. For example, the dots of the topmost star in Figure \ref{fig:umap}a show fingerprints of PHAs containing a nitro phenyl functional group, while the rightmost star includes fingerprints of PHAs containing benzonitrile. Similarly, the different clusters in Figure \ref{fig:umap}b have 13 corner vertices (some of them are hidden) that indicate the fingerprints of the 13 conventional polymers (see Figure \ref{fig:pha_design}a), which are included in the fingerprint subspace of Figure \ref{fig:umap}b, instead of the five PHAs as in Figure \ref{fig:umap}a. This agglomeration to stars or clusters illustrates that the used fingerprints (i) unambiguously distinguish polymers; (ii) position related polymers of similar chemistries in the vicinity (i.e., pack all copolymers with a specific side-chain functional group in the same part of the space); (iii) and thus create a well-conditioned and smooth learning problem well-suited for machine learning.



\begin{figure}[hbt]
 \includegraphics{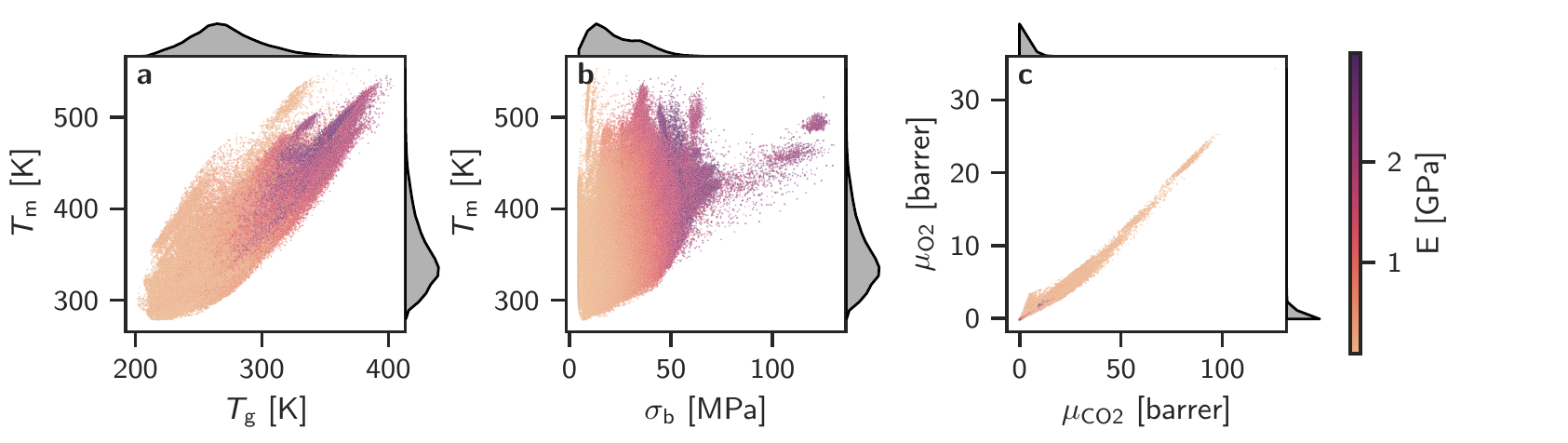}
  \caption{Property relations of almost 1.4 million bioplastic candidates. The data point densities are indicated in the plot margins. $T_\text{m}$, $T_\text{g}$, $\sigma_\text{b}$, $\mu_{\text{O}_2}$, $\mu_{\text{CO}_2}$, and E stand for melting temperature, glass transition temperature, tensile strength at break, O$_2$ gas permeability, CO$_2$ gas permeability, and Young's modulus, respectively.}
  \label{fig:scatter_plots}
\end{figure}

Figure \ref{fig:scatter_plots} displays property relations for a selected set of properties of the bioplastic search space in order to qualitatively assess our predictions and verify physical trends expected based on chemical intuition. The trend in Figure \ref{fig:scatter_plots}a is that polymers of high $T_\text{g}$ values also have high $T_\text{m}$ and room temperature $E$ values. This confirms our chemical intuition that $T_\text{g}$ is approximately linearly correlated to $T_\text{m}$ and high $T_\text{g}$ and/or $T_\text{m}$ polymers have stiffer morphologies thus possessing high $E$ values. Also, we observe that the correlation of $T_\text{g}$ and $T_\text{m}$ is not sharp but broad, which arises from the different side-chain functional groups in the search space. In contrast, Figure \ref{fig:scatter_plots}b suggests little to no correlation of $T_\text{m}$ and $\sigma_\text{b}$, except that the range spanned by $\sigma_\text{b}$ at a given $T_\text{m}$ broadens as $T_\text{m}$ increases. However, $\sigma_\text{b}$ is intuitively correlated to $E$, i.e., stiffer materials (high $E$) break at higher stresses (high $\sigma_\text{b}$). Figure \ref{fig:scatter_plots}c shows a roughly linear correlation of $\mu_{\text{CO}_2}$ and $\mu_{\text{O}_2}$  that again agrees with chemical intuition and lends credibility to the developed predictors. 

\paragraph{Bioplastic Replacements}

\begin{table}[hbtp]
  \caption{Measured properties of petroleum-based commodity plastics that in total account for \SI{75.1}{\%} of Europe's yearly plastic production in 2019 (see usage column)\cite{plasticseurop}. Property values are averaged over the entries in the PolyInfo repository \cite{polyinfo} at standard conditions.}
  \label{tbl:application} 

\resizebox{\textwidth}{!}{
\begin{tabular}{p{3cm}lp{4cm}p{3cm}rrrrrrrrr}
\hline 
 Polymer & Abb. & Applications & SMILES$^a$ & $T_\text{g}$  & $T_\text{m}$& $\sigma_\text{b}$ & $\epsilon_\text{b}$ & E  & $\mu_{\text{O}_2}$ & $\mu_{\text{CO}_2}$ & Usage \\
 
 &    &  &  & [K] & [K] & [MPa] & & [MPa] & [barrer] & [barrer] & [\%] \\
\hline
 Poly(ethylene) & PE & Cloth packaging, shopping bags, waste bags & \texttt{[*]CC[*]} & 220 & 403 & 22.0 & 338.0 &  670 & 2.00 & 27.10 & 29.2 \\
 Poly(propylene) & PP & Living hinges, pipes, caps, cutlery & \texttt{[*]CC([*])C} & 287 & 437 & 30.0 & 150.0 & 1600 & 0.76 & 4.40 & 19.3 \\
 Poly(vinyl chloride) & PVC & Window frames, cables, pipes, films & \texttt{[*]CC([*])Cl} & 353 & 485 & 36.0 & 29.7 & 1680 & 0.06 & 0.23 & 9.9 \\
 Poly(ethylene terephthalate) & PET & Bottles, automotive industry & \texttt{[*]CCOC(=O)c1c cc(C(=O)O[*])cc1} & 350 & 526 & 119.0 & 65.0 & 2970 & 0.05 & 0.33  & 7.9 \\
 Poly(styrene) & PS & Packaging fillers, cutlery, foam cups, take-out boxes & \texttt{[*]CC([*])c1 ccccc1}  & 371 & 528 & 34.0 & 2.0 & 2450 & 2.60 & 12.60 & 6.8 \\
 Poly(hexano-6-lactam) & Nylon6 & Yarns, fibers & \texttt{[*]CCCCCC (=O)N[*]} & 324 & 493 & 60.0 & 61.0 & 1600 & 4.00 & 0.09 & 2 \\
Poly(ethylene 2,6-naphthalate) & PEN & Bottles, scintillators, medical product containers & \texttt{[*]CCOC(=O)c1cc c2cc(C(=O)O[*]) ccc2c1} & 357 & 541 & 77.0 & 42.0 & 2310 & 0.02 & 0.24 & \\
\hline
\end{tabular}
}

\footnotesize{$^a$The two stars (\texttt{[*]}) indicate the endpoints of the polymer repeat unit.}
\end{table}

\begin{figure}[hbt]
 \includegraphics[width=1\textwidth]{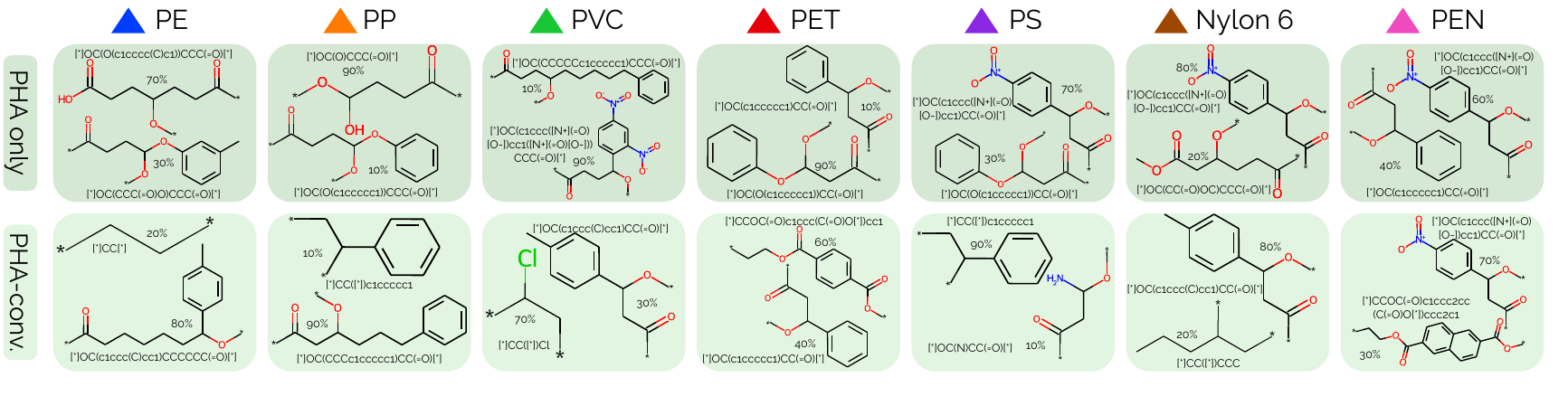}
  \caption{PHA-only and PHA-conventional bio-replacements for seven commodity plastics. Full polymer names are listed in Table \ref{tbl:application}.}
  \label{fig:biocandidates}
\end{figure}

Up to this point, we have discussed the training and validation of three multitask deep neural networks (each targeting separately the thermal, mechanical, and gas permeability properties) to forecast 13 polymer properties, the consideration of a search space of over 1.3 million bioplastic candidates, and predictions for each of the candidates in the search space. Next, we search the candidate set for suitable replacements for seven petroleum-based and commonly used plastics listed in Table \ref{tbl:application}. The search is performed following a two-step protocol. In the first step, we employ a nearest-neighbor search to find the five closest replacements (within the target property space) for each of the seven plastics and in each copolymer subgroup of PHAs-only and PHAs with conventional polymers. In the second step, we use our domain expertise to pick the most promising bio-replacement from the five candidates based on its synthesizability potential. The most promising bio-replacements for each commodity plastic and for each of the two copolymer subgroups are reported in Figure \ref{fig:biocandidates}. The full list of bio-replacements (70) is provided as a Supplementary Information TXT file.

\begin{figure}[hbt]
 \includegraphics{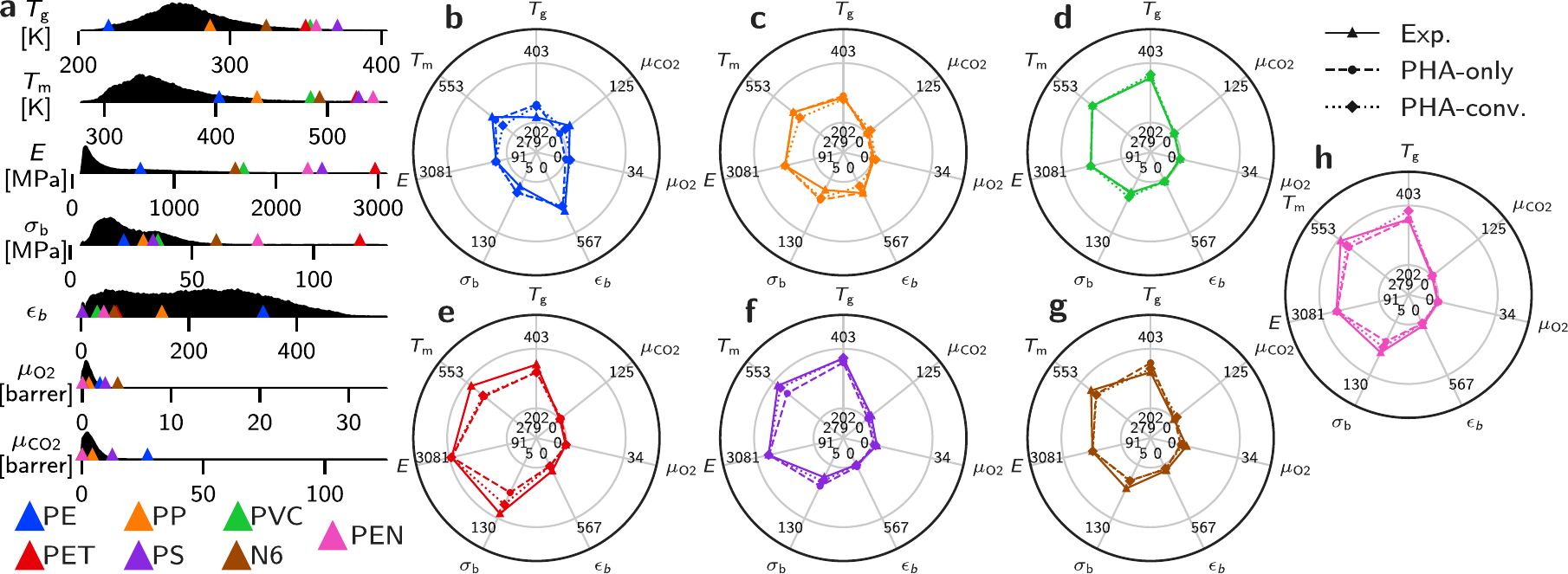}
  \caption{Experimental and predicted properties. \textbf{a} Property density profiles computed over the entire prediction set of bioplastic candidates. Missing x-axes beyond a certain cutoff indicate zero predicted property densities over those property ranges. The triangles show experimental properties of the seven commodity plastics. Full polymer names are listed in Table \ref{tbl:application}. \textbf{b} to \textbf{h} Property radar charts for each commodity plastics. Triangles with solid lines show the experimental properties. Circles with dashed lines and diamonds with dotted lines indicate predicted properties of the bio-replacements in Figure \ref{fig:biocandidates} for the copolymer subgroups of PHA-only and PHA-conventional polymers, respectively.}
  \label{fig:distributions}
\end{figure}

Figure \ref{fig:distributions}a shows the property distributions of the bioplastic candidate set along with the properties of the seven commodity plastics (c.f., Table \ref{tbl:application}) indicated as triangles. As expected, the $T_\text{m}$ peak is shifted to higher temperatures (by around \SI{80}{K}) with respect to the $T_\text{g}$ peak. Among the mechanical properties, the densities of $E$ and $\sigma_\text{b}$ demonstrate a peak at around \SI{540}{MPa} and \SI{20}{MPa}, while $\epsilon_\text{b}$ shows a broad distribution, which covers most of the data range. Moreover, the majority of $\mu_{\text{O}_2}$ and $\mu_{\text{CO}_2}$ values are below \SI{4}{barrer} and \SI{20}{barrer}, which match the expected value range of this polymer class \cite{polyinfo}. Overall, all commodity plastics (triangles) lie within the property ranges spanned by the bioplastic search space. However, because the triangles often lie in the tails of the property distributions, it is challenging (but possible) to find suitable replacements. Similar to Figure \ref{fig:distributions}a, Figures \ref{fig:distributions}b to \ref{fig:distributions}h compare the experimental properties of the commodity plastics with their top bio-replacements identified in Figure \ref{fig:biocandidates} in a radar chart. Qualitative graphical pairwise similarities between the property profiles in each radar chart indicate that the identified bio-replacements do indeed exhibit similar sets of properties with respect to the experimental properties. A comparison of the predicted and experimental properties of the seven commodity plastics is shown in Supplementary Figure S9. 



\paragraph{Synthesis Opportunities}
It is interesting to note that all PHA-only and PHA-conventional bio-replacements in Figure \ref{fig:biocandidates} contain aromatic groups in the side-chain. The biosynthesis of PHAs containing an aromatic monomer was first reported in 1990 by Fritzsche et al. \cite{fritzsche1990unusual} for Poly(3-hydroxy-5-phenylvalerate) and since then a wide range of aromatic side-chain functional groups have been introduced into PHAs through biosynthesis routes. For instance, see a recent comprehensive review by Ishii-Hyakutake et al. \cite{ishii2018biosynthesis} for a more detailed discussion. Looking at the complexity of the chemistries that are already accessible via biosynthesis, prospects of producing the identified PHA-only bio-replacements are rather optimistic. For example, Ar\'ostegui et al. \cite{Arostegui1999} reported that the \textit{Pseudomonas oleovorans} bacterium synthesizes PHAs with nitrophenyl side-chain functional groups, which occur in the PHA-only bio-replacements of PVC, PS, and PEN in Figure \ref{fig:biocandidates}. Moreover, engineering the bacterial PHA synthesis remains a highly active area of research with novel genome editing techniques, such as clustered regularly interspaced short palindromic repeats interference (CRISPRi),\cite{meng2017synthetic} that carry enormous potential for future breakthroughs in terms of both the accessible chemical diversity in PHA biosynthesis of homo- and copolymers as well as the yield optimization of the produced polymer chemistries.


Chemical synthesis routes for PHAs have been widely reported in the literature as well\cite{Westlie2020,Okada2002}. The potential for the chemical synthesis of the copolymers in this work (Figure \ref{fig:biocandidates} and Supplementary Information) lies in the ability to combine two comonomers of differing composition. Methods used in the synthesis of polystyrene-co-lactone copolymers\cite{Herman1981,Zalusky2002} can lead to the desired formation of the PHA-conventional bio-replacements of PP and PS. A chemical synthesis route for the PE and Nylon 6 PHA-conventional bio-replacements may follow similar steps used for the copolymerization of PHA/PEO (polyethylene oxide) copolymers\cite{Winnacker2017}. The PVC, PET and PEN bio-replacements from PHA-conventional polymers may be produced via a reactive twin-screw extrusion process to form block structures containing phthalate-co-lactones.\cite{Tang1999} Alternatively, a copolymer of repeating phthalate/lactone units has recently been produced via a copolymerization reaction of $\varepsilon$-lactone with degraded PET \cite{Espinoza2019,Ben2015}. We believe the predictions based on our work have potential to translate into new PHA biopolymers or copolymers and will inspire development of new PHA-only or hybrid conventional-PHA polymer synthesis routes.

\section{Conclusion}

We have developed an informatics-based bioplastic design pipeline, which has identified promising PHA-based bioplastic replacements for seven petroleum-based commodity plastics. Our study starts with the data collection and curation of approximately $23\,000$ homo- and copolymer data points spanning 13 properties critical for everyday applications and use. Multitask neural networks with a meta learner, pioneered by us for polymer informatics, forecast thermal, mechanical, and gas permeability properties for polymers over a broad chemical space with unprecedented performance.  Using the trained models, we predict the 13 key thermal, mechanical, and gas permeability properties of all polymers in a bioplastic search space of almost 1.4 million polymers. The property predictions are validated and subsequently utilized to find bio-replacements for seven commodity plastics that, in total, account for more than \SI{75}{\%} of the yearly plastic production. Using a two-step selection protocol of a nearest-neighbor search and synthesizability criteria, we propose two bio-replacements for each commodity plastic and discuss chemical synthesis and biosynthesis routes for these promising polymer replacements. Informatics can help to identify suitable synthesis strategies as well\cite{Chen2021}.

The implications of this work are far-reaching. We currently produce by far more plastics than we can recycle,\cite{Geyer2017} and the demand for plastics is expected to continue to grow at an annual rate of \SI{4}{\%}.\cite{rosenboom2022bioplastics} As countries begin to implement restrictions on plastic use, there is an urgent need for bioplastic alternatives to conventional plastics. Yet, the options of commercially available biopolymers are currently very limited.\cite{ghosh2021roadmap} Our approach to design and discover functional biopolymers can be applied to greatly accelerate the replacement of conventional plastic materials with more sustainable alternatives, and with possibly even greater performance advantages. The candidate biopolymers, in particular PHAs, might be synthesized by known chemical or biosynthetic routes, hybrid routes, or routes yet to be developed. Our approach can augment conventional empirically based design approaches by guiding the way to more targeted experiments, fewer experimental trials, or shorter times to market. Our work provides an informatics-based screening tool for researchers and developers aiming to produce bioplastics with improved thermomechanical and transport properties for better performance in specific applications, thus accelerating the transition to a circular economy.

\section{Methods}
\paragraph{Fingerprinting}

The fingerprinting process converts geometric and chemical information of polymers to machine-readable numerical representations for training machine learning models. Polymer structures are represented as simplified molecular-input line-entry system (SMILES)\cite{Weininger1988} strings that use two stars to indicate the two endpoints of the repetitive unit of the polymers, but otherwise follow the SMILES syntax. The fingerprint vector ($\mathbf{F}$) in this work has 849 components and is calculated based upon the SMILES string following a two-step protocol\cite{Kuenneth2021a}: First, we compute hierarchical fingerprints that capture structural and key chemical features of each comonomer at three different length scales\cite{Mannodi-Kanakkithodi2016}. At the atomic scale, our fingerprints track the occurrence of a fixed set of atomic fragments (or motifs)\cite{Huan2015}. For example, the fragment ‘‘C3-S2-C3’’ is composed of three contiguous atoms, in this order, a three-fold coordinated carbon, a two-fold coordinated sulfur, and a three-fold coordinated carbon. A vector of such triplets represents the fingerprint components at the lowest hierarchy. The next level uses the quantitative structure-property relationship (QSPR) fingerprints\cite{Le2012} to capture features on larger length-scales. QSPR fingerprints are often used in chemical and biological sciences, and used here as implemented in the chem informatics toolkit RDKit\cite{rdkit}. Examples of such fingerprints are the van der Waals surface area\cite{Iler1995}, the topological polar surface area (TPSA),\cite{Ertl2000,Prasanna2008} the fraction of atoms that are part of rings (i.e., the number of atoms associated with rings divided by the total number of atoms in the formula unit), and the fraction of rotatable bonds. The highest length-scale fingerprint components in our polymer fingerprints deal with ‘‘morphological descriptors’’. They include features such as the shortest topological distance between rings, the fraction of atoms that are part of side-chains, and the length of the largest side-chain. 

Second, we sum the composition-weighted comonomer fingerprints to compute the total copolymer fingerprint vector $\mathcal{F} = \sum_i^N \mathbf{F}_i c_i $, where $N$ is the number of comonomers in the copolymer, $\mathbf{F}_i$ the $i^{\text{th}}$ comonomer fingerprint, and $c_i$ the fraction of the $i^{\text{th}}$ comonomer. This copolymer fingerprint satisfies the two main demands of uniqueness and invariance to different (but equivalent) periodic unit specifications and renders the fingerprinting routine invariant to the order in which one may sort the comonomers. In our work, all copolymer data points are of random copolymers, and alternating copolymers were treated as homopolymers. 

\paragraph{Multitask Predictor and Meta Learner}

Multitask deep neural networks simultaneously learn multiple polymer properties to utilize possible inherent correlations in data. Figure \ref{fig:pha_design}b schematically portrays the architecture of the three concatenation-conditioned multitask predictors: the copolymer fingerprint and selector vector are fed to a feed-forward deep neural network that outputs a single property. The selector vector is a binary vector and specifies the output property. For instance, the selector vector of the thermal properties predictor ($S_1$) has three components and encodes $T_\text{g}$ as $[100]$, $T_\text{m}$ as $[010]$, and $T_\text{d}$ as $[001]$. All parameters of the neural networks, such as the number of layers, number of nodes, dropout rates, and activation functions, are optimized using the Hyperband method\cite{Li2018} of the Python package Keras-Tuner\cite{kerastuner}. Final parameters are reported in Supplementary Table S1. All models were implemented using the Python API of TensorFlow\cite{tensorflow}. 
 
The training protocol of the predictors follows state-of-the-art techniques involving five-fold cross-validation and a meta learner that forecasts the final property values based upon the ensemble of cross-validation predictors\cite{Kuenneth2021a}. The parameters of the cross-validation models are fixed when used in the meta learner. The meta learner has the same network architecture as the multitask predictors but receives the five outputs of the multitask predictors as inputs (rather than the copolymer fingerprint). The cross-validation process ensures that each data point has at least once been in the validation data set and allows us to report the generalization error as averaged RMSEs and $R^2$s of the validation data sets. The three meta learners operate as production predictors. After shuffling, the data set was split into two parts. \SI{20}{\%} of the data set was set aside for training the meta learners, while the remaining \SI{80}{\%} was used for five-fold cross-validation and the validation of the meta learner. All data set splits were stratified by the properties.

\section{Code Availability}
The code is available at \url{https://github.com/Ramprasad-Group/bioplastic_design} and production models are deployed openly at \url{https://PolymerGenome.org}.

\section{Competing Interests}
The authors declare no competing interests.

\section{Data Availability}
All the polymer data used in this work to train the various property prediction models can be found in the PolyInfo database https://polymer.nims.go.jp/en/ (National Institute for Materials Science (NIMS) holds the copyright of this database system). 


\begin{acknowledgement}

C.K. thanks the Alexander von Humboldt Foundation for financial support. This work is financially supported by the Office of Naval Research through a Multi-University Research Initiative (MURI) grant (N00014-17-1-2656) and a regular grant (N00014-20-2175). G.P., B.L.M. and C.N.I. acknowledge support from the Los Alamos National Laboratory (LANL) Laboratory Directed Research and Development (LDRD) program’s project titled Bio-Manufacturing with Intelligent Adaptive Control (BioManIAC) \#20190001DR. J.L. gratefully acknowledges support via a LANL Center for Nonlinear Studies (CNLS) Summer 2021 Fellowship Award. LANL is operated by Triad National Security, LLC, for the National Nuclear Security Administration of U.S. Department of Energy (Contract No. 89233218CNA000001).

\end{acknowledgement}

\section{Author Contributions}
C.K. designed, trained and evaluated the machine learning models. Numerous discussion with C.N.I. and B.L.M. at an early stage helped in defining the scope of the study. J.L. provided input for the background information and copolymer examples. The work was conceived and guided by R.R and G.P. All authors discussed the results and commented on the manuscript.

\begin{suppinfo}
The file \texttt{candidates.txt} (CSV format) contains 70 predicted bio-replacements for seven commodity plastics. The file \texttt{candidates.txt} is also available at
\url{https://github.com/Ramprasad-Group/bioplastic_design}. Supplementary Figures and Tables are available.

\end{suppinfo}

\clearpage
\bibliography{references}

\end{document}